\documentclass[10pt,letterpaper,twocolumn]{article}

\usepackage[margin=0.75in]{geometry}
\usepackage[T1]{fontenc}
\usepackage{newtxtext}
\usepackage[hyphens]{url}
\usepackage{graphicx}
\usepackage{amsmath,amssymb}
\usepackage{natbib}
\usepackage{booktabs}
\usepackage[hidelinks]{hyperref}
\title{APCReg: Anatomical-Prior-Guided Coarse-to-Fine CBCT--IOS Registration via Multi-View Projection and Reliability-Controlled Residual Correction}

\makeatletter
\renewcommand{\maketitle}{%
  \twocolumn[{%
    \begin{center}
      {\LARGE\bfseries \@title\par}
      \vspace{1.0em}

      \begin{minipage}{0.96\textwidth}
        \centering
        \large
        Xincan Zheng\textsuperscript{1},
        Yaqi Wang\textsuperscript{2,*},
        Zhi Li\textsuperscript{1},
        Jiahao Bao\textsuperscript{3},
        Lan Feng\textsuperscript{4},
        Yiru Xia\textsuperscript{5},
        Shuai Wang\textsuperscript{1,*}

        \vspace{0.7em}
        \small
        \textsuperscript{1}School of Cyberspace Security, Hangzhou Dianzi University, Hangzhou, China\\
        \textsuperscript{2}Innovation Center for Electronic Design Automation Technology, Hangzhou Dianzi University, Hangzhou, China\\
        \textsuperscript{3}Department of Craniomaxillofacial Surgery, Shanghai Ninth People's Hospital,\\
        Shanghai Jiao Tong University School of Medicine, Shanghai, China\\
        \textsuperscript{4}Department of Dentistry, Sir Run Run Shaw Hospital, Zhejiang University School of Medicine, Hangzhou, Zhejiang, China\\
        \textsuperscript{5}Department of Periodontology, Shanghai Stomatological Hospital \& School of Stomatology, Fudan University, Shanghai, China
      \end{minipage}
    \end{center}
    \vspace{1.0em}
  }]
}
\makeatother

\hypersetup{
  pdftitle={APCReg: Anatomical-Prior-Guided Coarse-to-Fine CBCT--IOS Registration via Multi-View Projection and Reliability-Controlled Residual Correction},
  pdfauthor={Xincan Zheng, Yaqi Wang, Zhi Li, Jiahao Bao, Lan Feng, Yiru Xia, Shuai Wang},
  pdfsubject={CBCT--IOS registration},
  pdfkeywords={CBCT, intraoral scan, point cloud registration, dental imaging, coarse-to-fine registration}
}

\begin{document}
\maketitle

\begingroup
\renewcommand{\thefootnote}{}
\footnotetext{\raggedright\textsuperscript{*}Corresponding authors: \nolinkurl{echowyq0154@gmail.com} (Yaqi Wang); \nolinkurl{shuaiwang.tai@gmail.com} (Shuai Wang).}
\addtocounter{footnote}{-1}
\endgroup

\begin{abstract}
Registration between cone-beam computed tomography (CBCT) and intraoral scans (IOS) is essential for patient-specific surgical planning. However, disparate imaging modalities, limited overlap, and large pose offsets make automated registration unreliable. Consequently, clinical registration remains dependent on conventional geometry pipelines and manual clinician adjustment. To address these challenges, we propose APCReg, an anatomical-prior-guided coarse-to-fine framework for global registration and reliability-controlled residual correction. Specifically, multi-view anatomical coarse registration (MACR) performs ordered orthogonal projection alignment (buccal, proximal, and occlusal) to decompose the six-degree-of-freedom search before three-dimensional refinement. Overlap-aware residual registration (OARR) combines shared KPConv features, a folded arch-length cue, overlap-gated cross-attention, and Sinkhorn matching. Finally, dental-arch-structured hypothesis selection evaluates diverse poses on held-out reliable correspondences, while a ground-truth-free coarse-retention guard conditionally retains a geometrically reliable coarse pose. On 60 held-out jaw pairs, APCReg achieves a submillimeter mean Chamfer distance of 0.87 mm and a Hausdorff distance of 2.92 mm under this evaluation protocol, and ranks first across the six reported metrics among the evaluated open-source baselines.
\end{abstract}

\begin{figure}[!t]
  \centering
  \includegraphics[width=0.93\columnwidth,
  keepaspectratio]{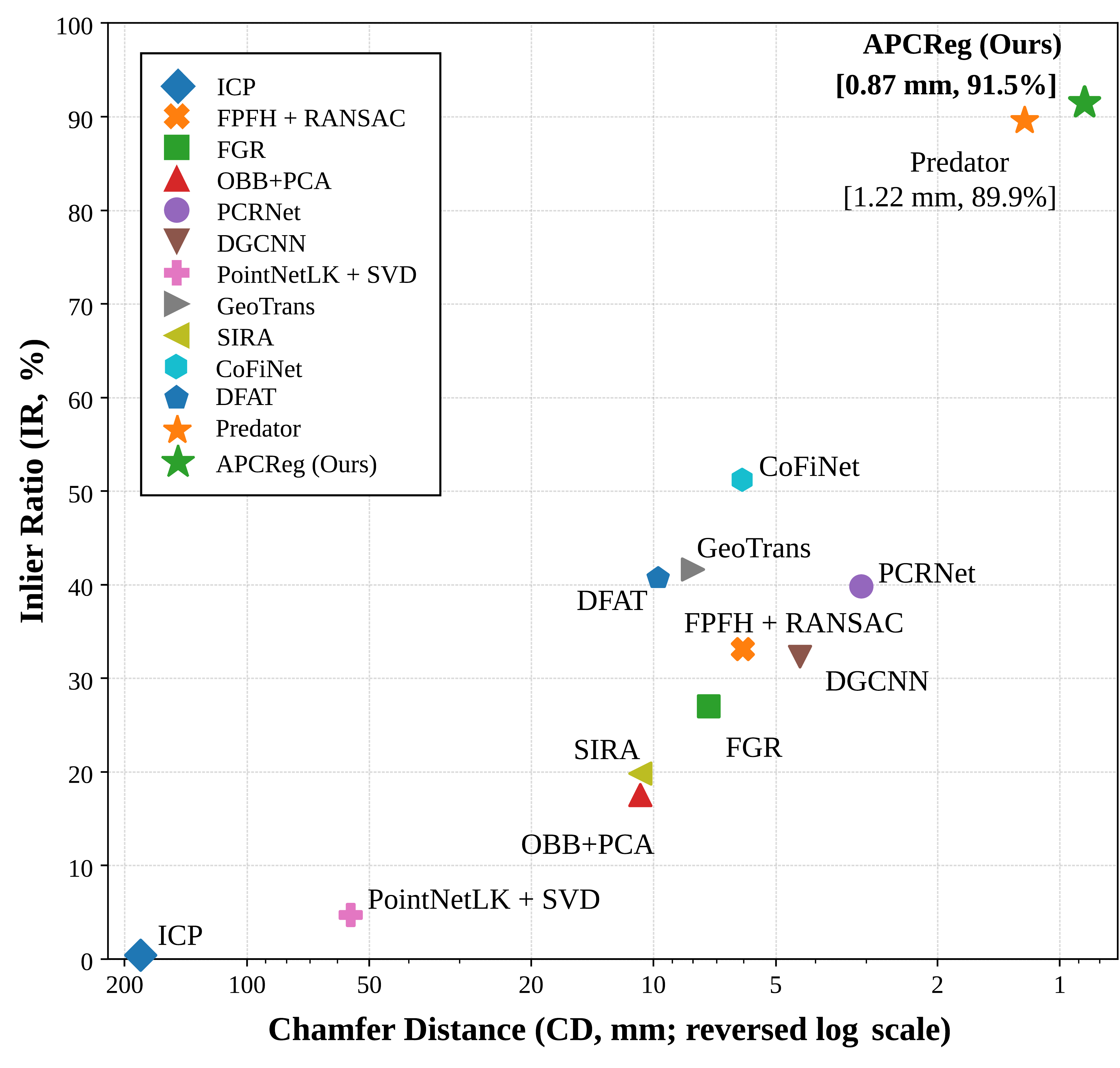}
    \caption{Test-set trade-off between Chamfer distance (CD; reversed logarithmic axis, lower is better) and inlier ratio (IR; higher is better). APCReg achieves $0.87\,\mathrm{mm}$ CD and $91.5\%$ IR.}
\label{fig:cd_ir_comparison}
\end{figure}
\begin{figure*}[!t]
  \centering
  \includegraphics[width=0.98\textwidth]{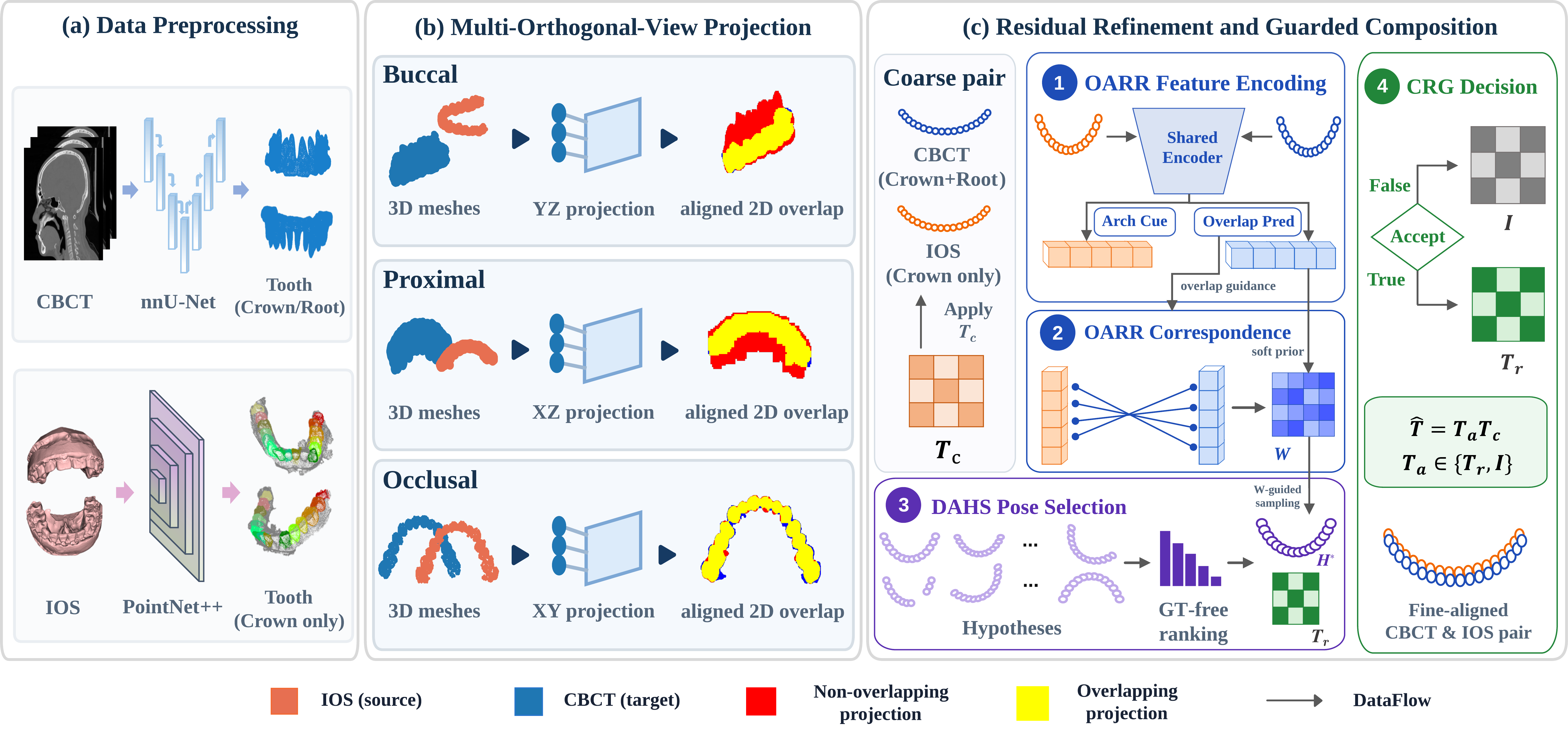}
    \caption{
    Overview of APCReg.    
    (a) nnU-Net and PointNet++ extract CBCT and IOS tooth surfaces. (b) MACR obtains the coarse transform $T_c$ through ordered buccal, proximal, and occlusal projection alignment. (c) OARR uses predicted overlap to guide matching and construct the soft transport matrix $W$; DAHS generates and ranks $W$-guided structured hypotheses, and refits the full transport under the selected hypothesis $H^{*}$ to obtain the residual transform $T_r$. CRG selects $T_a \in \{T_r,I\}$ before final composition, $\widehat{T}=T_aT_c$.    
    }
  \label{fig:overview}
\end{figure*}

\section{Introduction}

CBCT--IOS registration estimates a rigid transformation that maps an intraoral scan (IOS) into the cone-beam computed tomography (CBCT) frame, combining crown surfaces with root anatomy for digital treatment planning \citep{flugge2017cbctios,liu2023ddmf,kim2024fusion,zheng2025review}. Prior systems combine descriptors or landmarks with RANSAC/FGR and ICP, or use tooth-level global-to-local fitting \citep{chung2020automatic,kim2023curvature,jang2024cbctios,kim2024fusion}.

Accurate rigid registration requires reliable, distinctive cross-modal correspondences. Classical pipelines encode local geometry with handcrafted descriptors, estimate a global pose using RANSAC or FGR, and apply ICP for refinement \citep{rusu2009fpfh,zhou2016fgr,besl1992icp}. They are transparent and accurate once corresponding regions are proximal, but large rotations and translations remove the neighborhood evidence needed for matching and ICP convergence. Learning-based methods replace descriptors with learned features or direct pose regression, from PointNet-based PCRNet and PointNetLK to graph-based DGCNN encoders \citep{sarode2019pcrnet,aoki2019pointnetlk,wang2019dgcnn}. Correspondence-driven models learn descriptors, keypoints, confidence, and soft assignments \citep{zeng2017threedmatch,deng2018ppfnet,gojcic2019smoothnet,bai2020d3feat,wang2019dcp,choy2020dgr,yew2020rpmnet,bai2021pointdsc}; DeepVCP and 3DRegNet jointly predict correspondences and motion, while TEASER supplies certifiable robust estimation \citep{lu2019deepvcp,pais2020regnet,yang2021teaser}. Later architectures add overlap, superpoint, rigid-invariant, or domain-adaptive reasoning \citep{huang2021predator,yu2021cofinet,qin2022geotransformer,chen2023sira,fu2025dfat}. Nevertheless, cross-modal sampling and artifacts weaken transfer, and repeated crowns remain locally indistinguishable. Pairwise Euclidean objectives also fail to preserve dental-arch structure \citep{lin2026mci}: a one-tooth shift can look compatible while violating arch order. Coarse-to-fine pipelines separate global capture from local refinement \citep{yu2021cofinet,kim2024fusion}, but still require a coarse pose within the residual model's operating range; otherwise matching is poorly conditioned and an update may overwrite a valid initialization.

Consequently, reliable CBCT--IOS registration needs dental-arch structure beyond local geometry while retaining correspondence precision. APCReg addresses these gaps through successive decisions rather than monolithic pose prediction (Figure~\ref{fig:overview}). MACR recovers global pose through ordered buccal, proximal, and occlusal searches. OARR refines the residual using learned cross-modal geometry and a PCA-axis-reversal-invariant coordinate. DAHS evaluates directional, multiscale hypotheses with arch context, and CRG makes a ground-truth-free test-time decision between the selected update and identity.

Our contributions are threefold:
\begin{itemize}
  \item MACR decomposes the poorly conditioned six-degree-of-freedom global search into an ordered, principal-axis-aligned 2D silhouette-matching paradigm, providing robust initialization for anisotropic structures through explicit geometric priors.
  \item OARR and DAHS combine overlap-aware matching, a folded arch-length coordinate invariant to PCA-axis reversal, directional arch hypotheses, and held-out scoring to address ambiguous residual tooth shifts.
  \item CRG is a ground-truth-free test-time geometric retention mechanism that retains MACR whenever the residual violates a fixed geometric acceptance rule.
\end{itemize}

\section{Related Work}

\textbf{CBCT--IOS registration.}
CBCT--IOS fusion is commonly formulated as global-to-local geometric fitting \citep{zheng2025review}. Clinical workflows first established the need to align CBCT and intraoral surfaces for guided implant planning \citep{flugge2017cbctios}; later fully automatic systems combine projection/pose cues, clustered transformations, tooth segmentation, and stitching correction \citep{chung2020automatic,kim2023curvature,liu2023ddmf,jang2024cbctios,kim2024fusion}. A single global solve can still rely on unreliable local evidence before cross-modal overlap is localized. APCReg establishes overlap through view-wise recovery and reserves learned correspondence for the reduced residual regime.

\textbf{Geometric initialization and projection reasoning.}
Classical registration recovers a pose from geometric descriptors and robust consensus before local alignment. FPFH--RANSAC \citep{rusu2009fpfh}, FGR \citep{zhou2016fgr}, and ICP \citep{besl1992icp} remain attractive for their explicit correspondences, but descriptor matching and nearest-neighbor refinement require correct structures to be proximal. Projection-based dental registration supplies global silhouette evidence before this condition holds \citep{chung2020automatic}. MACR likewise assigns complementary degrees of freedom to ordered buccal, proximal, and occlusal views, then verifies anatomically plausible proposals before three-dimensional cleanup.

\textbf{Learned correspondence under partial overlap.}
Learning-based registration replaces descriptors with learned features, correspondence confidence, and differentiable pose recovery. PointNet-based PCRNet and graph-based DGCNN provide direct learned representations for pose or local structure \citep{sarode2019pcrnet,wang2019dgcnn}; DCP and RPM-Net estimate soft correspondences and recover rigid motion by differentiable alignment \citep{wang2019dcp,yew2020rpmnet}, whereas DGR and PointDSC learn correspondence confidence and spatial consistency \citep{choy2020dgr,bai2021pointdsc}. Predator, CoFiNet, and GeoTransformer model overlap, coarse-to-fine superpoint matches, or rigid-invariant context \citep{huang2021predator,yu2021cofinet,qin2022geotransformer}; SIRA and DFAT target domain shift \citep{chen2023sira,fu2025dfat}. Prior-guided experts \citep{huang2025psreg} and global context beyond Euclidean neighborhoods \citep{lin2026mci} further improve ambiguous-overlap matching. These advances strengthen local discrimination but do not encode the ordered dental-arch relation needed to reject a plausible one-tooth shift. OARR adds a PCA-axis-reversal-invariant arch coordinate, and DAHS evaluates directional hypotheses on held-out correspondences.

\textbf{Coarse-to-fine registration and pose acceptance.}
Coarse-to-fine designs separate wide-range capture from accurate residual fitting \citep{yu2021cofinet}. The MICCAI STSR 2025 challenge illustrates the difficulty of directly recovering large-offset CBCT--IOS poses: its first-ranked PointNetLK-style submission combines iterative global features with differentiable singular value decomposition \citep{aoki2019pointnetlk}, yet reports 46.47-mm translation and $165.30^\circ$ rotation errors \citep{wang2025stsr}. This motivates separating wide-range capture from residual refinement. Whereas existing pipelines generally accept the produced residual, APCReg separates estimation from deployment-time acceptance: OARR yields weighted correspondences, DAHS ranks structured candidates, and CRG compares the update with identity using geometric consistency, retaining MACR when the residual violates Eq.~\eqref{eq:guard_rule}.

\section{Method}

\subsection{Problem Formulation and Overview}

Given an IOS surface $\mathcal{S}=\{\mathbf{x}_i\}_{i=1}^{N}$ and a CBCT surface $\mathcal{Q}=\{\mathbf{y}_j\}_{j=1}^{M}$, APCReg estimates an orientation-preserving rigid transform represented as a homogeneous matrix $\mathbf{T}\in\mathrm{SE}(3)$, whose action on a point is $\mathbf{T}(\mathbf{x})=\mathbf{R}\mathbf{x}+\mathbf{t}$, where $\mathbf{R}\in\mathrm{SO}(3)$ and $\mathbf{t}\in\mathbb{R}^{3}$. Because IOS contains crowns whereas CBCT additionally contains roots, their overlap is unknown and asymmetric. APCReg first estimates a coarse transform $\mathbf{T}_c$ and then a residual $\mathbf{T}_r$ on $\mathcal{S}_c=\{\mathbf{T}_c(\mathbf{x}_i)\}_{i=1}^{N}$. A coarse-retention guard selects $\mathbf{T}_a\in\{\mathbf{T}_r,\mathbf{I}\}$, giving
\begin{equation}
 \widehat{\mathbf{T}}=\mathbf{T}_a\mathbf{T}_c.
 \label{eq:composition}
\end{equation}
With column vectors and right-to-left composition, $\mathbf{T}_a=\mathbf{I}$ retains the coarse pose. MACR captures global pose, OARR estimates residual correspondences, DAHS selects an anatomically consistent hypothesis, and CRG decides whether to apply it.

\subsection{Multi-View Anatomical Coarse Registration}

\begin{figure}[!t]
  \centering
  \includegraphics[width=0.95\columnwidth,
  keepaspectratio]{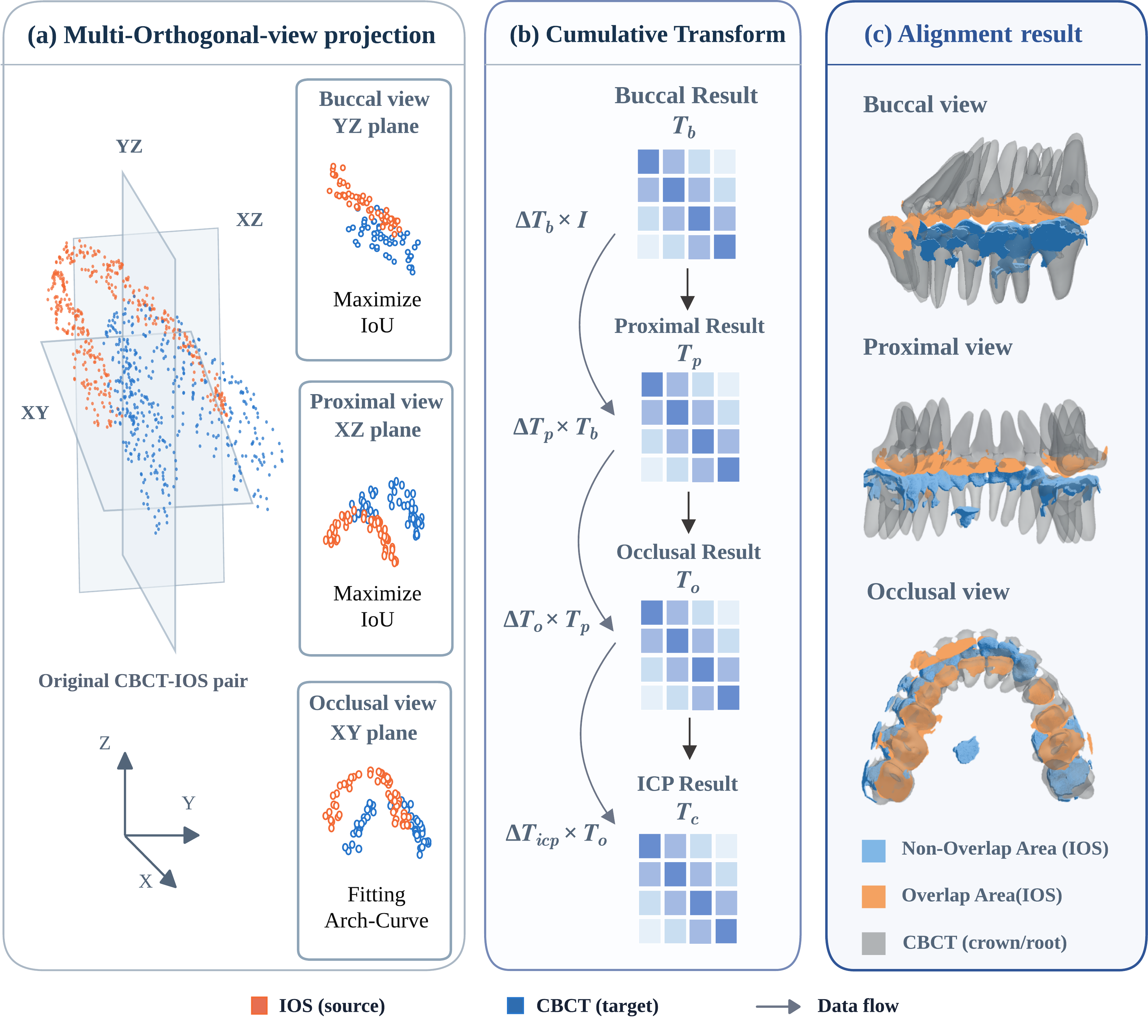}
    \caption{MACR coarse registration. \textbf{(a)} Buccal $YZ$, proximal $XZ$, and occlusal $XY$ projection alignment. \textbf{(b)} Sequential composition of view-specific updates followed by ICP yields $T_c$. \textbf{(c)} Resulting coarse alignment; gray denotes CBCT, while orange and light blue denote overlapping and non-overlapping IOS regions.}
  \label{fig:coarse}
\end{figure}

MACR replaces a poorly conditioned six-degree-of-freedom search with three ordered orthographic views. For $\mathbf{x}=(x,y,z)^\top$, the buccal, proximal, and occlusal projections are
\begin{equation}
 \Pi_b(\mathbf{x})=(y,z),\qquad
 \Pi_p(\mathbf{x})=(x,z),\qquad
 \Pi_o(\mathbf{x})=(x,y).
 \label{eq:view_projections}
\end{equation}
The buccal projection discards the $x$ coordinate and therefore optimizes only $(t_y,t_z,\theta_x)$. The proximal projection discards $y$ and optimizes $(t_x,t_z,\theta_y)$, whereas the occlusal projection discards $z$ and optimizes $(t_x,t_y,\theta_z)$. These are stage-specific increments rather than a simultaneous decomposition of one pose: a translation unobservable in one view is exposed by a later view, and repeated coordinates denote successive corrections in the fixed CBCT frame.

After anatomically signed OBB+PCA normalization, the longest dental extent defines $x$, the shortest extent approximating the occlusal normal defines $z$, and $y=z\times x$ completes a jaw-centered frame. The ordered $YZ$, $XZ$, and $XY$ views successively constrain jaw orientation and height, transverse displacement, and arch layout, yielding an anatomy-guided block-coordinate-style search. A candidate pose must therefore agree with complementary jaw profiles and the occlusal arch, rather than merely attaining high overlap in one projection.

A centroid-preserving $180^\circ$ flip and signed OBB+PCA alignment initialize the source. Both surfaces are rasterized in a shared metric frame, and each view maximizes silhouette IoU:
\begin{equation}
 J_v(\mathbf{T})=
 \frac{|\Omega_v^s(\mathbf{T})\cap\Omega_v^t|}
      {|\Omega_v^s(\mathbf{T})\cup\Omega_v^t|}.
 \label{eq:projection_iou}
\end{equation}
The buccal stage initializes projected-centroid displacement and optimizes $(t_y,t_z,\theta_x)$; the proximal stage initializes $t_x$ from bounding-box centers and refines $(t_x,t_z,\theta_y)$. The occlusal stage aligns extreme planes and robust quadratic arch curves in physical $XY$, accepting a directed partial-coverage proposal only when full-mask IoU does not decrease. Accepted increments are left-multiplied before local ICP yields $\mathbf{T}_c$.

\begin{figure}[!t]
  \centering
  \includegraphics[height=0.36\textheight,
  keepaspectratio]{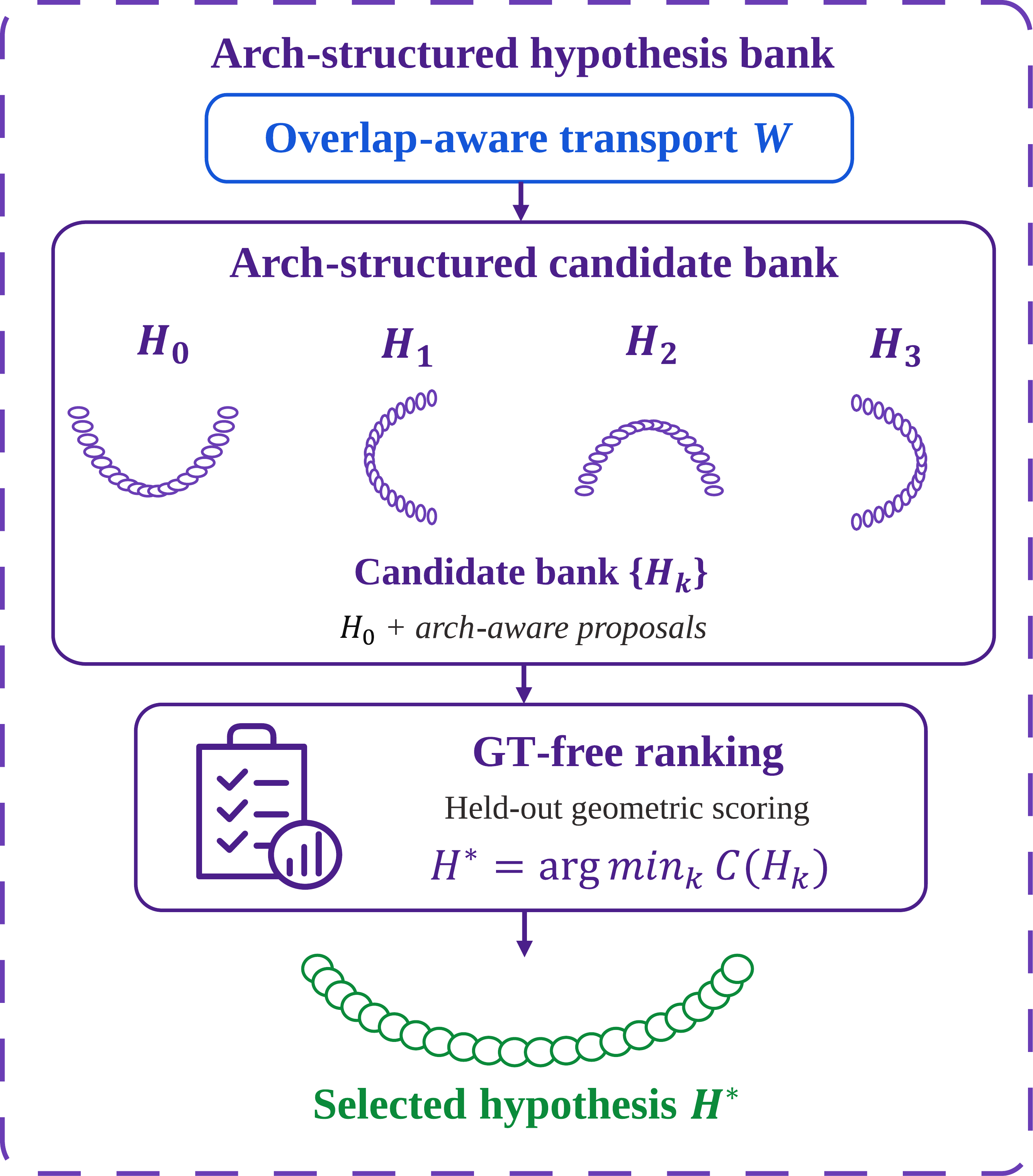}
      \caption{DAHS hypothesis selection. Overlap-aware transport generates an arch-structured candidate bank, and held-out, ground-truth-free geometric scoring selects the best hypothesis $\mathbf{H}^*$.
      }
  \label{fig:dahs}
\end{figure}

\subsection{Overlap-Aware Residual Registration}

OARR subtracts the shared pooled centroid $\mathbf{c}=(\sum_i\mathbf{x}_i^c+\sum_j\mathbf{y}_j)/(N+M)$, preserving physical scale and relative displacement, and conjugates the centered residual back to the CBCT frame.

After MACR, adjacent crowns may still have similar local geometry and CBCT-only regions may dominate unrestricted attention. OARR combines KPConv shape features, overlap probabilities, and a folded arch cue. Bottleneck points are sorted along the first three-dimensional principal direction, their normalized cumulative three-dimensional path length $u_i\in[0,1]$ is folded as $a_i=|2u_i-1|$, and the cue is disabled for degenerate spans. KPConv encodes local crown morphology, while the fold supplies a PCA-axis-reversal-invariant, non-injective arch-position cue; symmetric sites may share a value, so it is used as a positional cue rather than a correspondence label. Overlap prediction suppresses target-only regions before bidirectional cross-attention, and DAHS resolves the remaining directional ambiguity.

Overlap probabilities $q_i^s,q_j^t$ define the hard visibility sets
$I_s=\{i:q_i^s>0.3\}$ and $I_t=\{j:q_j^t>0.3\}$ for cross-attention.
Continuous overlap confidence gives

\begin{equation}
 \begin{aligned}
 W_{ij}&=\mathrm{stopgrad}(q_i^s q_j^t)A_{ij},\\
 &Z_i=\sum_j W_{ij},\
 \widetilde{\mathbf{y}}_i
 =\frac{\sum_j W_{ij}\bar{\mathbf{y}}_j}{Z_i}.
 \end{aligned}
 \label{eq:transport}
\end{equation}

Rows with $Z_i=0$ are discarded. The base centered residual is the determinant-corrected weighted Procrustes solution
\begin{equation}
 (\bar{\mathbf{R}},\bar{\mathbf{t}})\in
 \arg\min_{\mathbf{R}\in\mathrm{SO}(3),\,\mathbf{t}\in\mathbb{R}^3}
 \sum_{i:Z_i>0}Z_i
 \|\mathbf{R}\bar{\mathbf{x}}_i+\mathbf{t}
 -\widetilde{\mathbf{y}}_i\|_2^2,
 \label{eq:weighted_procrustes}
\end{equation}
which defines $\bar{\mathbf{T}}_0$. Fewer than three supported rows or cross-covariance rank below two skips residual refinement.

\subsection{Dental-Arch-Structured Hypothesis Selection}

Soft transport may be confident yet globally wrong because neighboring crowns can have similar features. DAHS filters correspondences using overlap, transport entropy, keypoint spacing, and transport mass; constructs normalized source and target arch coordinates $s_i^s,s_j^t\in[0,1]$; and evaluates competing poses on correspondences disjoint from those used for fitting (Figure~\ref{fig:dahs}). Because either principal-plane basis may be reflected, both source directions are enumerated. For direction $d\in\{+1,-1\}$, scale $\zeta>0$, and shift $b$, the arch gate is
\begin{equation}
 G_{ij}^{d,\zeta,b}=
 \exp\!\left[
 -\frac{\{s_j^t-[\zeta(s_i^{s,d}-\tfrac12)+\tfrac12+b]\}^2}
 {2\sigma_s^2}\right],
 \label{eq:arch_gate}
\end{equation}
where $s_i^{s,+}=s_i^s$ and $s_i^{s,-}=1-s_i^s$.

The bank contains soft, hard, arch-gated, visibility-balanced, and fixed-seed random transport variants. Each variant is fitted on $F$ to produce a pose hypothesis $\mathbf{H}_k\in\mathrm{SE}(3)$, which is evaluated on a disjoint, region-balanced set $E$. Its score combines trimmed residual, worst-region error, between-region variation, many-to-one collision, insufficient arch span, and a bounded learned correction:
\begin{equation}
 \begin{split}
 C_k={}&C_{\mathrm{trim}}(\mathbf{H}_k)
 +\lambda_w\max_r C_r(\mathbf{H}_k)
 +\lambda_b\,\mathrm{std}_r C_r(\mathbf{H}_k)\\
 &+\lambda_c C_{\mathrm{col}}(\mathbf{H}_k)
 +\lambda_s C_{\mathrm{span}}(\mathbf{H}_k)
 +\eta\tanh f_\phi(\boldsymbol{\varphi}_k),
 \end{split}
 \label{eq:hypothesis_score}
\end{equation}
The trimmed term summarizes supported residuals, whereas the worst-region and variation terms discourage uneven agreement across arch segments. Collision and span penalties reject many-to-one transport and hypotheses that explain only a short arch interval. A candidate must therefore satisfy both local residual and cross-region structural criteria.
Here $\boldsymbol{\varphi}_k\in\mathbb{R}^{12}$ collects geometric and support statistics for a LayerNorm MLP. Since $|\tanh(\cdot)|\leq1$, learning changes the geometric score by at most $\eta$. After selecting $\mathbf{H}^*=\arg\min_k C_k$, Gaussian pose consistency gates the full transport and weighted Procrustes is recomputed; invalid arch geometry returns $\bar{\mathbf{T}}_0$, whereas a rank-deficient final fit returns the selected valid hypothesis. The disjoint, region-balanced evaluation prevents self-scoring and density domination.

\subsection{Coarse-Retention Guard}

CRG compares residual refinement with identity using one-sided IOS-to-CBCT distances, avoiding a penalty for CBCT-only roots. Let $d_{(1)}(\mathbf{T})\leq\cdots\leq d_{(n)}(\mathbf{T})$ order $d_i(\mathbf{T})=\min_{\mathbf{y}\in\mathcal{Q}}\|\mathbf{T}(\mathbf{x}_i^c)-\mathbf{y}\|_2$. With $k=\lceil\alpha n\rceil$, define $g(\mathbf{T})=k^{-1}\sum_{r=1}^{k}d_{(r)}(\mathbf{T})+\beta d_{(\lceil\tau n\rceil)}(\mathbf{T})$. The first term summarizes the closest supported fraction of IOS points, whereas the percentile term retains sensitivity to broader residual error. The score therefore favors consistent crown alignment without requiring CBCT-only roots to be explained. For $0<\alpha,\tau\leq1$, $\beta,\delta,\gamma\geq0$, CRG returns
\begin{equation}
 \mathbf{T}_a=
 \begin{cases}
 \mathbf{I},&
 g(\mathbf{T}_r)>g(\mathbf{I})+\delta
 \ \mathrm{and}\ g(\mathbf{I})\leq\gamma,\\
 \mathbf{T}_r,&\mathrm{otherwise}.
 \end{cases}
 \label{eq:guard_rule}
\end{equation}
Identity is selected only when the residual is worse by margin $\delta$ and the coarse pose satisfies $\gamma$. We set $(\alpha,\tau,\beta,\gamma,\delta)=(0.80,0.90,0.25,1.30\,\mathrm{mm},0.15\,\mathrm{mm})$ on validation and freeze all values for testing.

\begin{table}[!t]
\centering
\small
\setlength{\tabcolsep}{1.5pt}
\begin{tabular}{@{}lrrrrrr@{}}
\toprule
Method & CD$\downarrow$ & RTE$\downarrow$ & RRE$\downarrow$ & RMSE$\downarrow$ & HD$\downarrow$ & IR$\uparrow$ \\
\midrule
\multicolumn{7}{c}{\textit{Classical geometric registration}} \\
\midrule
ICP & 182.54 & 72.21 & 166.20 & 183.13 & 213.91 & 0.4 \\
FPFH+RANSAC & 6.03 & 205.62 & 79.27 & 7.41 & 19.72 & 33.2 \\
FGR & 7.31 & 274.02 & 93.38 & 9.21 & 22.56 & 27.0 \\
OBB+PCA & 10.76 & 352.31 & 130.65 & 12.85 & 31.12 & 17.5 \\
\midrule
\multicolumn{7}{c}{\textit{Direct learned pose estimation}} \\
\midrule
PCRNet & 3.08 & 40.39 & 12.16 & 3.64 & 10.09 & 39.8 \\
DGCNN & 4.36 & 59.44 & 16.38 & 5.25 & 14.50 & 32.3 \\
PointNetLK+SVD & 55.56 & 372.84 & 149.79 & 57.88 & 83.33 & 4.7 \\
\midrule
\multicolumn{7}{c}{\textit{Learned correspondence and overlap reasoning}} \\
\midrule
GeoTrans & 8.00 & 146.74 & 61.32 & 9.88 & 23.66 & 41.6 \\
SIRA & 10.76 & 315.69 & 126.41 & 12.76 & 34.23 & 19.8 \\
CoFiNet & 6.05 & 225.78 & 68.39 & 7.34 & 18.20 & 51.2 \\
DFAT & 9.73 & 166.71 & 57.01 & 11.28 & 25.30 & 40.7 \\
Predator & \underline{1.22} & \underline{29.18} & \underline{6.94} & \underline{1.40} & \underline{3.90} & \underline{89.9} \\
\midrule
\multicolumn{7}{c}{\textit{Proposed method}} \\
\midrule
\textbf{APCReg} & \textbf{0.87} & \textbf{13.10} & \textbf{3.88} & \textbf{1.00} & \textbf{2.92} & \textbf{91.5} \\
\bottomrule
\end{tabular}
\caption{Comparison with open-source registration baselines on 60 held-out jaw pairs. Methods are grouped by registration paradigm.}
\label{tab:main}
\end{table}

\section{Experiments}
\subsection{Data, Protocol, and Metrics}
\textbf{Dataset and split.}
We use the MICCAI STSR 2025 CBCT--IOS benchmark \citep{wang2025stsr}. The benchmark contains 479 cases, including 179 pose-labeled cases available under its controlled-access protocol. Both IOS and CBCT use the same left-handed coordinate convention; therefore, a valid rigid mapping between them should satisfy $\det(\mathbf R)=+1$. We excluded 48 labels with $\det(\mathbf R)=-1$, as they contain a reflection and cannot be represented in $\mathrm{SE}(3)$. The remaining 131 cases were divided into 81/20/30 cases for training/validation/testing, with both jaws kept in the same partition; each case contains upper/lower jaw pairs.

IOS meshes are tooth-segmented using PointNet++ \citep{qi2017pointnetpp} pretrained on Teeth3DS \citep{ben2022teeth3ds,ben2023dteethseg}, and CBCT volumes using the nnU-Net \citep{isensee2021nnunet} plug-in in 3D Slicer \citep{fedorov2012slicer}. APCReg and all baselines receive only these tooth surfaces.

\textbf{Comparison protocol.}
All methods share surfaces, splits, transform direction, 10,000-point Poisson-disk samples, and evaluation code, and return a finite transform for every test pair. All baseline implementations are based on publicly available GitHub projects; trainable baselines use the same training split, with tunable choices selected on validation and no test-time tuning. Paired CD comparisons use 20,000 case-clustered bootstrap resamples (seed 20260716), preserving within-case jaw dependence.

\textbf{Metrics.}
We evaluate both transforms on 10,000 Poisson-disk-sampled IOS points, excluding CBCT-only roots. CD averages the two directed nearest-neighbor distances; HD is their maximum, and RMSE uses the predicted-IOS-to-ground-truth-IOS direction. RRE is the geodesic rotation error and RTE is the norm of the translation component of the relative transform. IR is the fraction of GT-overlapping IOS points whose predictions lie within $2\,\mathrm{mm}$ of the ground-truth IOS surface. Lower is better except for IR; distances are in millimeters and RRE in degrees.

\subsection{Comparison with Registration Baselines}

\begin{figure}[!t]
  \centering
  \includegraphics[height=0.31\textheight,
    keepaspectratio]{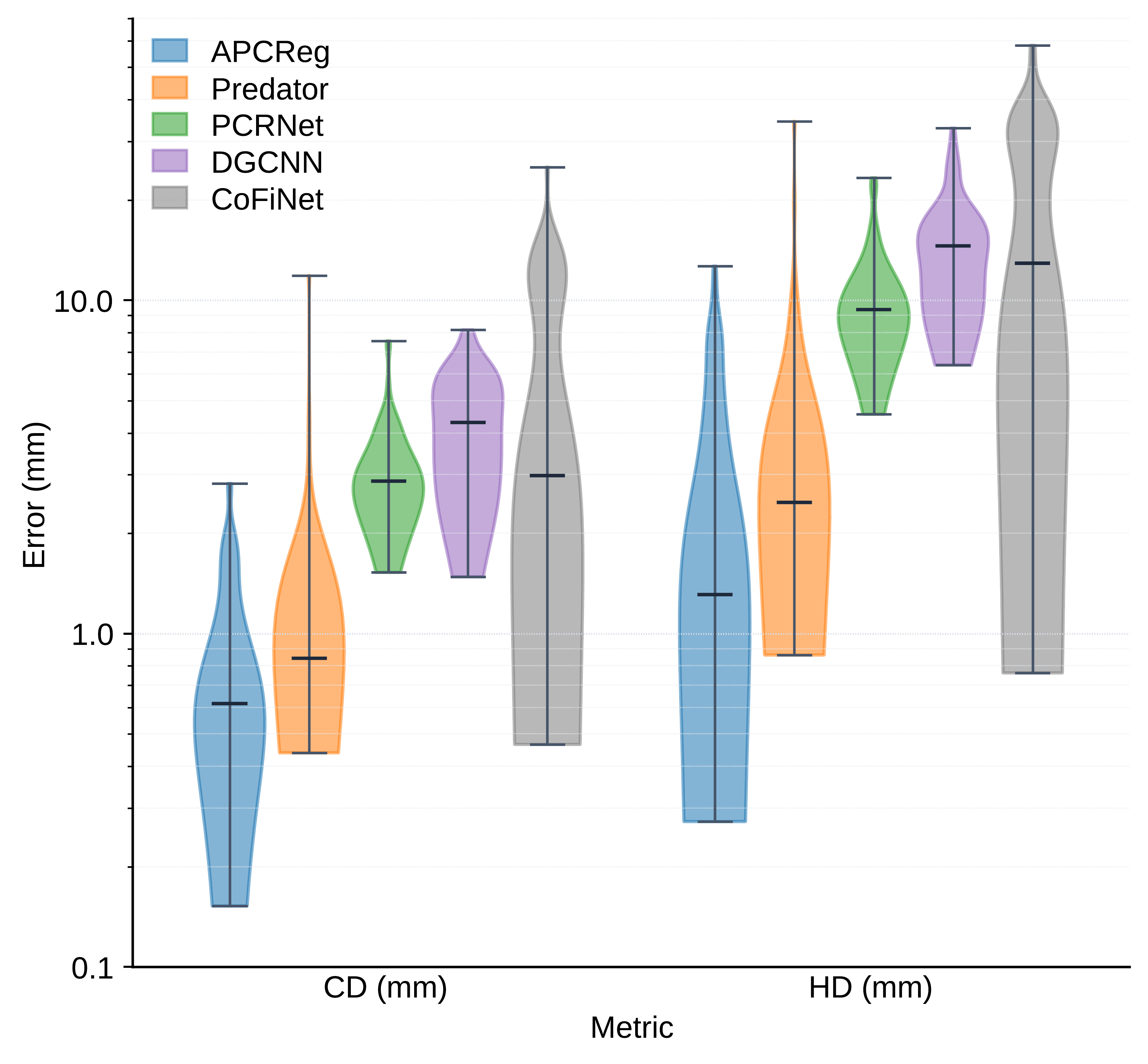}
    \caption{CD and HD distributions of five selected high-performing methods on 60 held-out jaw pairs. The vertical axis is logarithmic. Violin plots show the per-pair distributions; horizontal bars indicate medians, capped vertical bars indicate the observed ranges.
    }
  \label{fig:metric-distribution}
\end{figure}

\begin{figure*}[!t]
  \centering
  \includegraphics[width=\textwidth, 
    keepaspectratio]{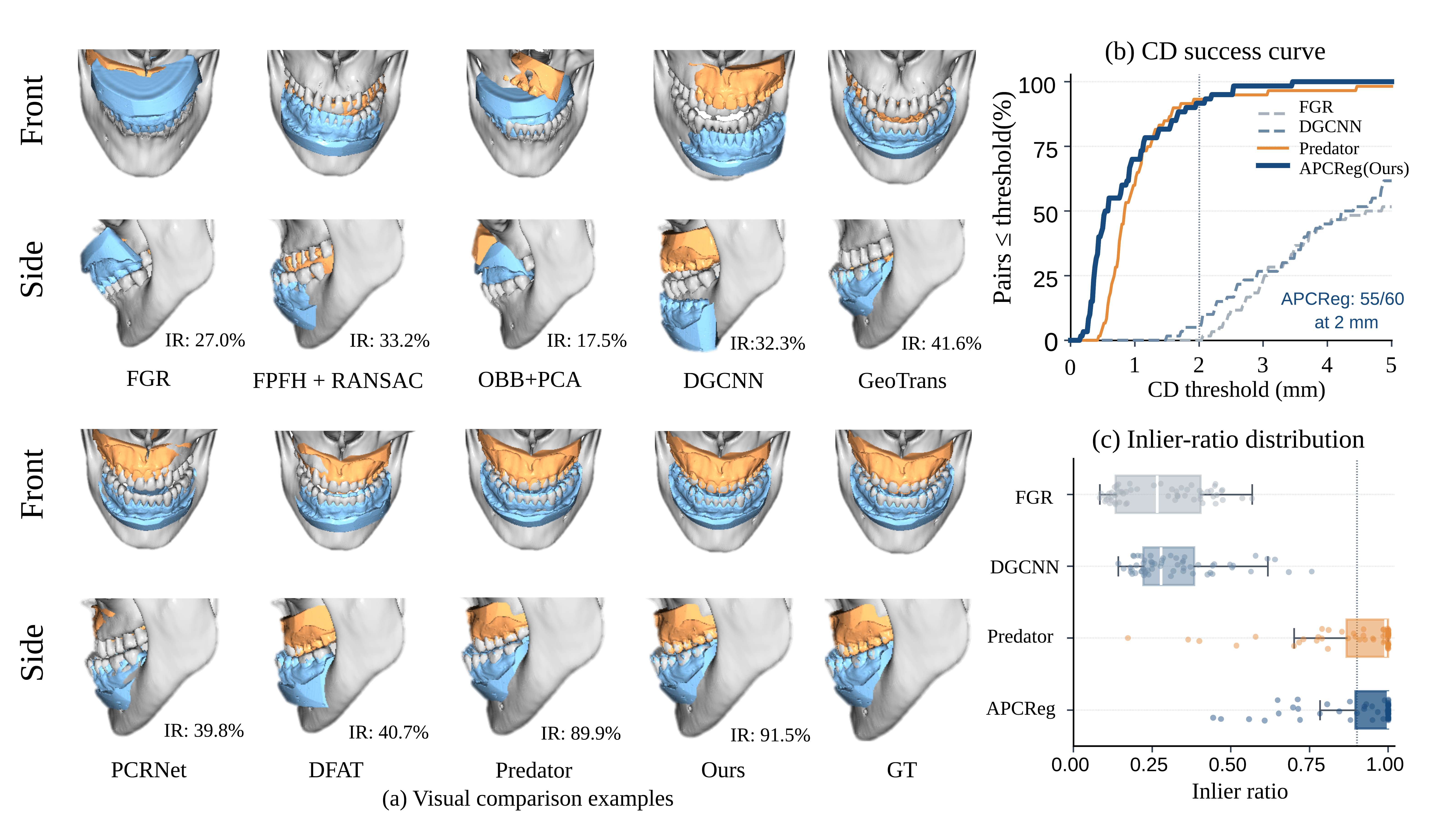}
    \caption{Registration comparison. \textbf{(a)} Frontal and lateral overlays for selected methods and ground truth; gray denotes CBCT, while orange and blue denote transformed upper- and lower-jaw IOS surfaces. \textbf{(b)} CD success curves, with APCReg registering $55/60$ pairs within $2\,\mathrm{mm}$. \textbf{(c)} Per-pair IR distributions; boxes indicate medians and interquartile ranges.
    }
  \label{fig:qualitative}
\end{figure*}

The baseline panel covers classical geometry, direct learned pose estimation (PCRNet \citep{sarode2019pcrnet} and DGCNN \citep{wang2019dgcnn}), differentiable PointNetLK, and learned correspondence/overlap methods. APCReg ranks first on all six metrics in Table~\ref{tab:main} among the evaluated open-source baselines. Relative to Predator, the strongest baseline, it reduces CD from 1.22 to 0.87 mm (28.4\%), RRE from $6.94^\circ$ to $3.88^\circ$, and RTE from 29.18 to 13.10 mm, while increasing IR from $89.9\%$ to $91.5\%$. The CD gain comprises 40/60 pairwise wins and a 0.345-mm mean reduction (95\% clustered-bootstrap CI: $[0.043,0.791]$).

\textbf{Top-method error distributions.}
Figure~\ref{fig:metric-distribution} compares the per-jaw CD and HD distributions of five selected high-performing methods. APCReg has the lowest median CD and HD, ahead of Predator, the strongest baseline. The remaining methods have higher central errors and broader tails. In Figure~\ref{fig:qualitative}, FGR and DGCNN retain large displacements and Predator leaves tooth-level offsets, whereas APCReg more consistently recovers the reference crown alignment.

\begin{table}[!ht]
\centering
\small
\setlength{\tabcolsep}{3.5pt}
\begin{tabular}{@{}lrrrr@{}}
\toprule
MACR stage & CD & RTE & RRE & IR \\
\midrule
Buccal & 78.15 & 115.53 & 12.64 & 5.4 \\
$+$ Proximal & 2.70 & 38.66 & 12.69 & 46.4 \\
$+$ Occlusal & 2.19 & 38.54 & 12.76 & 59.7 \\
B--P--O $+$ ICP & \textbf{1.28} & \textbf{14.51} & \textbf{5.22} & \textbf{82.8} \\
\midrule
Disrupted P--O--B $+$ ICP & 2.05 & \underline{19.09} & \underline{6.91} & \underline{67.4} \\
Reversed O--P--B $+$ ICP & \underline{2.03} & 19.66 & 7.08 & 66.0 \\
\bottomrule
\end{tabular}
\caption{Stage-wise MACR ablation and order stress test on 60 held-out jaw pairs.}
\label{tab:coarse}
\end{table}

\begin{table}[!ht]
\centering
\small
\setlength{\tabcolsep}{3.0pt}
\begin{tabular}{@{}lrrrr@{}}
\toprule
Residual variant & CD & RTE & RRE & IR \\
\midrule
A: Encoder + Sinkhorn & 1.65 & 16.41 & \underline{4.62} & 71.4 \\
A+B: + Cross-attention & 1.57 & \underline{16.33} & 4.65 & 71.7 \\
A+B+C: + Random Hypotheses & \underline{1.16} & 19.51 & 5.59 & 82.3 \\
A+B+C+D: + Structured & 1.18 & 17.95 & 5.13 & \underline{83.5} \\
A+B+C+D+E: + Learned scorer & 1.29 & 18.10 & 5.28 & 79.6 \\
\midrule
Full APCReg & \textbf{0.87} & \textbf{13.10} & \textbf{3.88} & \textbf{91.5} \\
\bottomrule
\end{tabular}
\caption{Cumulative APCReg ablation and final test-time geometric retention on 60 held-out jaw pairs.}
\label{tab:residual}
\end{table}
\subsection{Mechanism Validation}

\textbf{MACR stage contributions and order robustness.}
In Table~\ref{tab:coarse}, proximal and occlusal alignment substantially reduce CD, after which ICP yields the best performance. Reordering the same stages increases CD, with paired penalties of 0.769~mm (95\% CI: $[0.472,1.068]$) for P--O--B and 0.749~mm (95\% CI: $[0.525,0.986]$) for O--P--B. The number of pairs above 2-mm CD rises from 15 to 26 and 29, respectively. These results support the proposed anatomical schedule: stabilize orientation, recover the translation not observable in the buccal silhouette, and resolve arch layout before local refinement.

\textbf{Residual hypothesis reasoning and guarded selection.}
All variants in Table~\ref{tab:residual} share MACR initialization. Random hypotheses improve CD, whereas structured hypotheses yield the best unguarded IR. Together, these results show that candidate diversity and explicit arch structure provide complementary residual gains. The bounded learned correction remains subordinate to the geometric score when candidate costs are close. CRG then compares the selected residual with the identity alternative and retains MACR using inference geometry alone, further reducing CD and raising IR.

\textbf{CRG audit and deployment paths.}

\begin{figure}[!ht]
  \centering
  \includegraphics[width=\columnwidth]{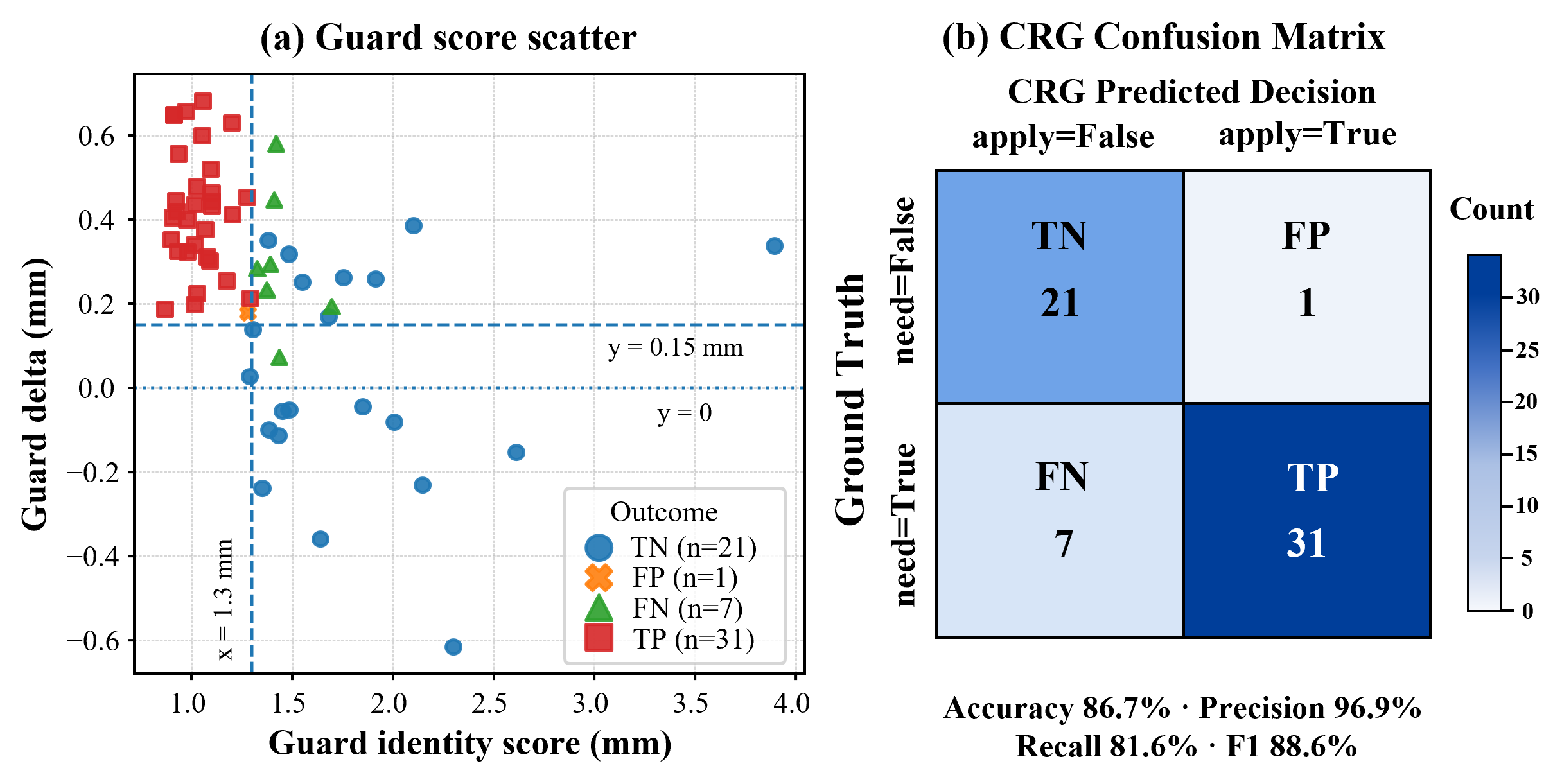}
    \caption{CRG audit on 60 held-out pairs. \textbf{(a)} Identity score versus $\Delta s=s_{\mathrm{pred}}-s_{\mathrm{id}}$; dashed thresholds at $1.30$ and $0.15\,\mathrm{mm}$ define the upper-left fallback region. \textbf{(b)} Fallback decisions against the ground-truth audit labels.}
\label{fig:crg-audit}
\end{figure}

The audit distinguishes harmful refinement from registration failure. For audit only, a refinement is harmful when $\mathit{need\_gt}=\mathbf{1}\!\left[\operatorname{CD}_{\mathcal{S}}(T_rT_c,T_{\mathrm{gt}})>\operatorname{CD}_{\mathcal{S}}(T_c,T_{\mathrm{gt}})\right]$, where $\operatorname{CD}_{\mathcal{S}}$ is the same-source CD computed on the sampled IOS surface. Raw refinement lowers mean CD from 1.389 to 1.157 mm and improves the 2-mm success rate from 39/60 to 56/60 (Table~\ref{tab:crg-audit}). CRG correctly retains the coarse pose for 31 of 38 harmful refinements, further reducing mean CD to 0.869 mm while preserving a 55/60 success count. All guard hyperparameters were selected on the validation set and frozen before test evaluation.

\begin{table}[!ht]
\centering
\small
\setlength{\tabcolsep}{3.5pt}
\begin{tabular}{@{}lrr@{}}
\toprule
Deployment path & Mean audit CD  & CD$\leq$2 mm \\
\midrule
Coarse only & 1.389 & 39/60 (65.0\%) \\
Raw refinement (no CRG) & \underline{1.157} & \textbf{56/60 (93.3\%)} \\
Refinement + CRG & \textbf{0.869} & \underline{55/60 (91.7\%)} \\
\bottomrule
\end{tabular}
\caption{Three deployment paths in the CRG audit on 60 held-out jaw pairs.}
\label{tab:crg-audit}
\end{table}

\textbf{Hard-case evaluation.}
Before baseline comparison, we define the hard subset as all 17 test jaw pairs whose MACR initialization has either $\mathrm{RRE}\geq10^\circ$ or $\mathrm{RTE}\geq20\,\mathrm{mm}$; all pairs meeting either condition are included.

\begin{table}[!t]
\centering
\small
\setlength{\tabcolsep}{2pt}
\begin{tabular}{@{}lrrrrr@{}}
\toprule
Method & CD & RMSE & HD & IR & CD$\leq$2mm \\
\midrule
ICP & 202.292 & 202.942 & 213.91 & 0.0 & 0/17 (0.0\%) \\
FPFH+RANSAC & 6.837 & 8.347 & 23.056 & 30.8 & 1/17 (5.9\%) \\
FGR & 7.584 & 9.589 & 23.835 & 24.7 & 0/17 (0.0\%) \\
OBB+PCA & 10.332 & 12.465 & 29.101 & 20.0 & 0/17 (0.0\%) \\
PCRNet & 3.365 & 3.953 & 11.389 & 36.1 & 0/17 (0.0\%) \\
DGCNN & 4.707 & 5.651 & 15.284 & 29.8 & 1/17 (5.9\%) \\
PointNetLK+SVD & 120.822 & 123.011 & 146.804 & 4.6 & 0/17 (0.0\%) \\
GeoTrans & 8.128 & 9.916 & 25.366 & 34.4 & 3/17 (17.6\%) \\
SIRA & 11.109 & 11.928 & 34.937 & 15.2 & 0/17 (0.0\%) \\
DFAT & 5.863 & 6.830 & 19.934 & 41.9 & 3/17 (17.6\%) \\
CoFiNet & 5.357 & 6.437 & 15.534 & 54.1 & 8/17 (47.1\%) \\
Predator & \textbf{1.233} & \textbf{1.419} & \textbf{4.544} & \textbf{87.8} & \textbf{15/17 (88.2\%)} \\
APCReg & \underline{1.594} & \underline{1.898} & \underline{6.655} & \underline{74.8} & \underline{14/17 (82.4\%)} \\
\bottomrule
\end{tabular}
\caption{Comparison on 17 hard jaw pairs with RTE and RRE omitted for compactness.}
\label{tab:hard}
\end{table}

Full APCReg reduces MACR's mean CD from 2.876 to 1.594 mm (44.6\%) and raises the 2-mm success count from 3/17 to 14/17, recovering 11 coarse failures. It therefore recovers most difficult initializations and remains competitive with the strongest baseline on this deliberately challenging subset.

\paragraph{Failure modes and scope.}
The hard-subset results in Table ~\ref{tab:hard} delineate APCReg's residual operating regime: even when MACR leaves large rotation or translation errors, APCReg recovers 11 coarse failures. CRG further reduces mean error by rejecting harmful residuals through a fixed geometric retention rule, without requiring calibrated uncertainty or test-time ground truth. Cases with substantial disagreement between the coarse and residual poses remain appropriate candidates for manual review. 

\section{Conclusion}

APCReg integrates projection-constrained global recovery, arch-aware residual matching, structured pose hypotheses, and test-time geometric retention. On 60 held-out jaw pairs, it ranks first across all six metrics among the evaluated open-source baselines, achieving 0.87-mm CD and $91.5\%$ IR. Order controls are consistent with MACR's buccal--proximal--occlusal schedule, while CRG conditionally retains a reliable coarse pose when the residual violates its acceptance rule. By assigning global recovery to MACR and learned matching to the residual regime, APCReg aligns each stage with the available anatomical evidence for CBCT--IOS registration. A natural next step is prospective, multi-center validation across scanners and dentitions, together with clinician-in-the-loop evaluation.

\bibliographystyle{plainnat}
\bibliography{ref}

@inproceedings{aoki2019pointnetlk,
  author    = {Aoki, Yasuhiro and Goforth, Hunter and Srivatsan, Rangaprasad Arun and Lucey, Simon},
  title     = {{PointNetLK}: Robust and Efficient Point Cloud Registration Using {PointNet}},
  booktitle = {Proceedings of the IEEE/CVF Conference on Computer Vision and Pattern Recognition},
  pages     = {7163--7172},
  year      = {2019}
}

@inproceedings{bai2021pointdsc,
  author    = {Bai, Xuyang and Luo, Zixin and Zhou, Lei and Chen, Hongkai and Li, Liu and Hu, Zhijian and Fu, Hongbo and Tai, Chiew-Lan},
  title     = {{PointDSC}: Robust Point Cloud Registration Using Deep Spatial Consistency},
  booktitle = {Proceedings of the IEEE/CVF Conference on Computer Vision and Pattern Recognition},
  pages     = {15859--15869},
  year      = {2021}
}

@article{ben2022teeth3ds,
  author  = {Ben-Hamadou, Anas and Smaoui, Omar and Chaabouni-Chouayakh, Houda and Rekik, Asma and Pujades, Sergi and Boyer, Edmond and Strippoli, Julien and Thollot, Antoine and Setbon, Hadrien and Trosset, Claire and Ladroit, Emmanuel},
  title   = {{Teeth3DS}: A Benchmark for Teeth Segmentation and Labeling from Intra-Oral {3D} Scans},
  journal = {arXiv preprint arXiv:2210.06094},
  year    = {2022},
  url     = {https://arxiv.org/abs/2210.06094}
}

@article{ben2023dteethseg,
  author  = {Ben-Hamadou, Anas and Smaoui, Omar and Rekik, Asma and Pujades, Sergi and Boyer, Edmond and Lim, Hyun-Kyung and Kim, Minseo and Lee, Minju and Chung, Minwoo and Shin, Young-Gil and Leclercq, Maxime and Cevidanes, Lucia and Prieto, Juan C. and Zhuang, Shuo and Wei, Guanglei and Cui, Zhiqiang and Zhou, Yifan and Dascalu, Teodor and Ibragimov, Bulat and Yong, Tien-Hui and Ahn, Hyun-Gi and Kim, Wooram and Han, Jae-Hoon and Choi, Byungdo and van Nistelrooij, Niels and Kempers, Sebastiaan and Vinayahalingam, Sharan and Strippoli, Julien and Thollot, Antoine and Setbon, Hadrien and Trosset, Claire and Ladroit, Emmanuel},
  title   = {{3DTeethSeg}'22: {3D} Teeth Scan Segmentation and Labeling Challenge},
  journal = {arXiv preprint arXiv:2305.18277},
  year    = {2023},
  url     = {https://arxiv.org/abs/2305.18277}
}

@article{besl1992icp,
  author  = {Besl, Paul J. and McKay, Neil D.},
  title   = {A Method for Registration of 3-D Shapes},
  journal = {IEEE Transactions on Pattern Analysis and Machine Intelligence},
  volume  = {14},
  number  = {2},
  pages   = {239--256},
  year    = {1992},
  doi     = {10.1109/34.121791}
}

@inproceedings{chen2023sira,
  author    = {Chen, Shiqi and Xu, Haobo and Li, Ruixin and Liu, Guofeng and Fu, Chi-Wing and Liu, Shuaicheng},
  title     = {{SIRA-PCR}: Sim-to-Real Adaptation for {3D} Point Cloud Registration},
  booktitle = {Proceedings of the IEEE/CVF International Conference on Computer Vision},
  pages     = {14394--14405},
  year      = {2023}
}

@inproceedings{choy2020dgr,
  author    = {Choy, Christopher and Dong, Wei and Koltun, Vladlen},
  title     = {Deep Global Registration},
  booktitle = {Proceedings of the IEEE/CVF Conference on Computer Vision and Pattern Recognition},
  pages     = {2514--2523},
  year      = {2020}
}

@article{chung2020automatic,
  author  = {Chung, Minyoung and Lee, Jingyu and Song, Wisoo and Song, Youngchan and Yang, Il-Hyung and Lee, Jeongjin and Shin, Yeong-Gil},
  title   = {Automatic Registration between Dental Cone-Beam {CT} and Scanned Surface via Deep Pose Regression Neural Networks and Clustered Similarities},
  journal = {IEEE Transactions on Medical Imaging},
  volume  = {39},
  number  = {12},
  pages   = {3900--3909},
  year    = {2020},
  doi     = {10.1109/TMI.2020.3007520}
}

@inproceedings{fu2025dfat,
  author    = {Fu, Kexue and Yuan, Ming and Wang, Cheng and Pang, Wenjie and Chi, Junhao and Wang, Ming and Gao, Liang},
  title     = {Dual Focus-Attention Transformer for Robust Point Cloud Registration},
  booktitle = {Proceedings of the IEEE/CVF Conference on Computer Vision and Pattern Recognition},
  pages     = {11769--11778},
  year      = {2025}
}

@article{fedorov2012slicer,
  author  = {Fedorov, Andriy and Beichel, Reinhard and Kalpathy-Cramer, Jayashree and Finet, Julien and Fillion-Robin, Jean-Christophe and Pujol, Sonia and Bauer, Christian and Jennings, Dominique and Fennessy, Fiona M. and Sonka, Milan and Buatti, John and Aylward, Stephen R. and Miller, James V. and Pieper, Steve and Kikinis, Ron},
  title   = {{3D Slicer} as an Image Computing Platform for the Quantitative Imaging Network},
  journal = {Magnetic Resonance Imaging},
  volume  = {30},
  number  = {9},
  pages   = {1323--1341},
  year    = {2012},
  doi     = {10.1016/j.mri.2012.05.001}
}

@inproceedings{huang2021predator,
  author    = {Huang, Shengyu and Gojcic, Zan and Usvyatsov, Mikhail and Wieser, Andreas and Schindler, Konrad},
  title     = {Predator: Registration of {3D} Point Clouds with Low Overlap},
  booktitle = {Proceedings of the IEEE/CVF Conference on Computer Vision and Pattern Recognition},
  pages     = {4267--4276},
  year      = {2021}
}

@inproceedings{huang2025psreg,
  author    = {Huang, Xiaoshui and Huang, Zhou and Zuo, Yifan and Gong, Yongshun and Zhang, Chengdong and Liu, Deyang and Fang, Yuming},
  title     = {{PSReg}: Prior-guided Sparse Mixture of Experts for Point Cloud Registration},
  booktitle = {Proceedings of the AAAI Conference on Artificial Intelligence},
  volume    = {39},
  pages     = {3788--3796},
  year      = {2025}
}

@article{isensee2021nnunet,
  author  = {Isensee, Fabian and Jaeger, Paul F. and Kohl, Simon A. A. and Petersen, Jens and Maier-Hein, Klaus H.},
  title   = {{nnU-Net}: A Self-Configuring Method for Deep Learning-Based Biomedical Image Segmentation},
  journal = {Nature Methods},
  volume  = {18},
  number  = {2},
  pages   = {203--211},
  year    = {2021},
  doi     = {10.1038/s41592-020-01008-z}
}

@article{kim2023curvature,
  author  = {Kim, Minchang and Chung, Minyoung and Shin, Yeong-Gil and Kim, Bohyoung},
  title   = {Automatic Registration of Dental {CT} and {3D} Scanned Model Using Deep Split Jaw and Surface Curvature},
  journal = {Computer Methods and Programs in Biomedicine},
  volume  = {233},
  pages   = {107467},
  year    = {2023},
  doi     = {10.1016/j.cmpb.2023.107467}
}

@inproceedings{kim2024fusion,
  author    = {Kim, Seonghyeon and Choi, Yejin and Na, Jaeseok and Song, In-Seok and Lee, Young-Seok and Hwang, Byung-Yoon and Lim, Hyun-Kyung and Baek, Seung Jun},
  title     = {Best of Both Modalities: Fusing {CBCT} and Intraoral Scan Data into a Single Tooth Image},
  booktitle = {Medical Image Computing and Computer Assisted Intervention--MICCAI 2024},
  series    = {Lecture Notes in Computer Science},
  volume    = {15002},
  pages     = {553--563},
  year      = {2024}
}

@article{liu2023ddmf,
  author  = {Liu, Jingwei and Hao, Jian and Lin, Haoyu and Pan, Wei and Yang, Jun and Feng, Yuxuan and Wang, Guorui and Li, Jiaxin and Jin, Zhen and Zhao, Zhi and Liu, Zhijian},
  title   = {Deep Learning-Enabled {3D} Multimodal Fusion of Cone-Beam {CT} and Intraoral Mesh Scans for Clinically Applicable Tooth--Bone Reconstruction},
  journal = {Patterns},
  volume  = {4},
  number  = {11},
  pages   = {100825},
  year    = {2023},
  doi     = {10.1016/j.patter.2023.100825}
}

@inproceedings{lin2026mci,
  author    = {Lin, Shuyuan and Peng, Wenwu and Huang, Junjie and Qi, Qiang and Wang, Miaohui and Weng, Jian},
  title     = {{MCI-Net}: A Robust Multi-Domain Context Integration Network for Point Cloud Registration},
  booktitle = {Proceedings of the AAAI Conference on Artificial Intelligence},
  volume    = {40},
  pages     = {23585--23593},
  year      = {2026}
}

@inproceedings{qi2017pointnetpp,
  author    = {Qi, Charles R. and Yi, Li and Su, Hao and Guibas, Leonidas J.},
  title     = {{PointNet++}: Deep Hierarchical Feature Learning on Point Sets in a Metric Space},
  booktitle = {Advances in Neural Information Processing Systems},
  volume    = {30},
  pages     = {5099--5108},
  year      = {2017}
}

@inproceedings{qin2022geotransformer,
  author    = {Qin, Zhe and Yu, Hao and Wang, Changjiang and Guo, Yulan and Peng, Yuxin and Xu, Kai},
  title     = {{GeoTransformer}: Fast and Robust Point Cloud Registration with Geometric Transformer},
  booktitle = {Proceedings of the IEEE/CVF Conference on Computer Vision and Pattern Recognition},
  pages     = {11143--11152},
  year      = {2022}
}

@inproceedings{rusu2009fpfh,
  author    = {Rusu, Radu Bogdan and Blodow, Nico and Beetz, Michael},
  title     = {Fast Point Feature Histograms ({FPFH}) for {3D} Registration},
  booktitle = {Proceedings of the IEEE International Conference on Robotics and Automation},
  pages     = {3212--3217},
  year      = {2009},
  doi       = {10.1109/ROBOT.2009.5152473}
}

@article{sarode2019pcrnet,
  author  = {Sarode, Vishwanath and Li, Xiang and Goforth, Hunter and Aoki, Yasuhiro and Srivatsan, Rangaprasad Arun and Lucey, Simon and Choset, Howie},
  title   = {{PCRNet}: Point Cloud Registration Network Using {PointNet} Encoding},
  journal = {arXiv preprint arXiv:1908.07906},
  year    = {2019},
  url     = {https://arxiv.org/abs/1908.07906}
}

@article{wang2019dgcnn,
  author  = {Wang, Yue and Sun, Yongbin and Liu, Ziwei and Sarma, Sanjay E. and Bronstein, Michael M. and Solomon, Justin M.},
  title   = {Dynamic Graph {CNN} for Learning on Point Clouds},
  journal = {ACM Transactions on Graphics},
  volume  = {38},
  number  = {5},
  pages   = {146:1--146:12},
  year    = {2019},
  doi     = {10.1145/3326362}
}

@article{wang2025stsr,
  author  = {Wang, Yifan and Li, Zeyu and Wu, Cheng and Liu, Jiaqi and Zhang, Yu and Chen, Jing and Ni, Jia and Luo, Qixuan and Liu, Jie and Han, Chen and Ji, Chao and Tan, Zhi Qiang and George, Adithya B. and Chen, Lei and Zhang, Qian and Qian, Dan and Wang, Shuang and Zhou, Hong},
  title   = {{MICCAI} {STSR} 2025 Challenge: Semi-Supervised Teeth and Pulp Segmentation and {CBCT}--{IOS} Registration},
  journal = {arXiv preprint arXiv:2512.02867},
  year    = {2025},
  url     = {https://arxiv.org/abs/2512.02867}
}

@inproceedings{wang2019dcp,
  author    = {Wang, Yue and Solomon, Justin M.},
  title     = {Deep Closest Point: Learning Representations for Point Cloud Registration},
  booktitle = {Proceedings of the IEEE/CVF International Conference on Computer Vision},
  pages     = {3523--3532},
  year      = {2019}
}

@inproceedings{yu2021cofinet,
  author    = {Yu, Hao and Li, Fu and Saleh, Mohamed and Busam, Benjamin and Ilic, Slobodan},
  title     = {{CoFiNet}: Reliable Coarse-to-Fine Correspondences for Robust Point-Cloud Registration},
  booktitle = {Advances in Neural Information Processing Systems},
  volume    = {34},
  pages     = {23872--23884},
  year      = {2021}
}

@inproceedings{yew2020rpmnet,
  author    = {Yew, Zi Jian and Lee, Gim Hee},
  title     = {{RPM-Net}: Robust Point Matching Using Learned Features},
  booktitle = {Proceedings of the IEEE/CVF Conference on Computer Vision and Pattern Recognition},
  pages     = {11824--11833},
  year      = {2020}
}

@article{zheng2025review,
  author  = {Zheng, Qianhan and Wu, Yongjia and Chen, Jiahao and Wang, Xiaozhe and Zhou, Mengqi and Li, Huimin and Lin, Jiaqi and Zhang, Weifang and Chen, Xuepeng},
  title   = {Automatic Multimodal Registration of Cone-Beam Computed Tomography and Intraoral Scans: A Systematic Review and Meta-Analysis},
  journal = {Clinical Oral Investigations},
  volume  = {29},
  pages   = {97},
  year    = {2025},
  number  = {2},
  doi     = {10.1007/s00784-025-06183-x}
}

@inproceedings{zhou2016fgr,
  author    = {Zhou, Qian-Yi and Park, Jaesik and Koltun, Vladlen},
  title     = {Fast Global Registration of 3D Point Clouds},
  booktitle = {European Conference on Computer Vision},
  pages     = {766--782},
  year      = {2016}
}

@inproceedings{zeng2017threedmatch,
  author    = {Zeng, Andy and Song, Shuran and Niessner, Matthias and Fisher, Matthew and Xiao, Jianxiong and Funkhouser, Thomas},
  title     = {{3DMatch}: Learning Local Geometric Descriptors from {RGB-D} Reconstructions},
  booktitle = {Proceedings of the IEEE Conference on Computer Vision and Pattern Recognition},
  pages     = {1802--1811},
  year      = {2017}
}

@inproceedings{deng2018ppfnet,
  author    = {Deng, Haowen and Birdal, Tolga and Ilic, Slobodan},
  title     = {{PPFNet}: Global Context Aware Local Features for Robust {3D} Point Matching},
  booktitle = {Proceedings of the IEEE Conference on Computer Vision and Pattern Recognition},
  pages     = {195--205},
  year      = {2018},
  doi       = {10.1109/CVPR.2018.00028}
}

@inproceedings{gojcic2019smoothnet,
  author    = {Gojcic, Zan and Zhou, Caifa and Wegner, Jan D. and Wieser, Andreas},
  title     = {The Perfect Match: {3D} Point Cloud Matching with Smoothed Densities},
  booktitle = {Proceedings of the IEEE/CVF Conference on Computer Vision and Pattern Recognition},
  pages     = {5545--5554},
  year      = {2019},
  doi       = {10.1109/CVPR.2019.00569}
}

@inproceedings{lu2019deepvcp,
  author    = {Lu, Weixin and Wan, Guowei and Zhou, Yao and Fu, Xiangyu and Yuan, Pengfei and Song, Shiyu},
  title     = {{DeepVCP}: An End-to-End Deep Neural Network for Point Cloud Registration},
  booktitle = {Proceedings of the IEEE/CVF International Conference on Computer Vision},
  pages     = {12--21},
  year      = {2019}
}

@inproceedings{pais2020regnet,
  author    = {Pais, G. Dias and Ramalingam, Srikumar and Govindu, Venu Madhav and Nascimento, Jacinto C. and Chellappa, Rama and Miraldo, Pedro},
  title     = {{3DRegNet}: A Deep Neural Network for {3D} Point Registration},
  booktitle = {Proceedings of the IEEE/CVF Conference on Computer Vision and Pattern Recognition},
  pages     = {7193--7203},
  year      = {2020}
}

@inproceedings{bai2020d3feat,
  author    = {Bai, Xuyang and Luo, Zixin and Zhou, Lei and Fu, Hongbo and Quan, Long and Tai, Chiew-Lan},
  title     = {{D3Feat}: Joint Learning of Dense Detection and Description of {3D} Local Features},
  booktitle = {Proceedings of the IEEE/CVF Conference on Computer Vision and Pattern Recognition},
  pages     = {6359--6367},
  year      = {2020},
  doi       = {10.1109/CVPR42600.2020.00639}
}

@article{yang2021teaser,
  author  = {Yang, Heng and Shi, Jingnan and Carlone, Luca},
  title   = {{TEASER}: Fast and Certifiable Point Cloud Registration},
  journal = {IEEE Transactions on Robotics},
  volume  = {37},
  number  = {2},
  pages   = {314--333},
  year    = {2021},
  doi     = {10.1109/TRO.2020.3033695}
}

@article{flugge2017cbctios,
  author  = {Fl{\"u}gge, Tabea and Derksen, Wiebe and te Poel, Jobine and Hassan, Bassam and Nelson, Katja and Wismeijer, Daniel},
  title   = {Registration of Cone-Beam Computed Tomography Data and Intraoral Surface Scans---A Prerequisite for Guided Implant Surgery with {CAD}/{CAM} Drilling Guides},
  journal = {Clinical Oral Implants Research},
  volume  = {28},
  number  = {9},
  pages   = {1113--1118},
  year    = {2017},
  doi     = {10.1111/clr.12925}
}

@article{jang2024cbctios,
  author  = {Jang, Tae Jun and Yun, Hye Sun and Hyun, Chang Min and Kim, Jong-Eun and Lee, Sang-Hwy and Seo, Jin Keun},
  title   = {Fully Automatic Integration of Dental {CBCT} Images and Full-Arch Intraoral Impressions with Stitching Error Correction via Individual Tooth Segmentation and Identification},
  journal = {Medical Image Analysis},
  volume  = {93},
  pages   = {103096},
  year    = {2024},
  doi     = {10.1016/j.media.2024.103096}
}

\end{document}